\documentclass[1p,times]{elsarticle}
\usepackage{graphicx}
\usepackage{subfigure}
\usepackage{amssymb}
\usepackage{amsmath}
\usepackage{bm}

\def\simleq{\; \raise0.3ex\hbox{$<$\kern-0.75em \raise-1.1ex\hbox{$\sim$}}\; }
\def\simgeq{\; \raise0.3ex\hbox{$>$\kern-0.75em \raise-1.1ex\hbox{$\sim$}}\; }

\newcommand{\GeV}{{\rm GeV}}

\newcommand{\TeV}{{\rm TeV}}

\begin{document}

\title{A consistent interpretation of recent CR nuclei and electron spectra}

\author{Giuseppe Di Bernardo}
\address{Department of Physics, University of Gothenburg, SE-412 96, Gothenburg, Sweden}
\author{Carmelo Evoli} 
\address{SISSA, via Bonomea 265, 34136 Trieste, Italy}
\author{Daniele Gaggero, \underline{Dario Grasso}\footnote{\tt dario.grasso@pi.infn.it}}
\address{INFN, Sezione di Pisa, Largo Bruno Pontecorvo 3, 56127 Pisa, Italy}
\author{Luca Maccione}
\address{Deutsches Elektronen-Synchrotron, Notkestra{\ss}e 85, 22607
Hamburg, Germany}
\author{Mario Nicola Mazziotta}
\address{Istituto Nazionale di Fisica Nucleare, Sezione di Bari, 70126 Bari, Italy}



\begin{abstract}
We try to interpret the recently updated measurement of the cosmic ray electron (CRE) spectrum observed by Fermi-LAT, together with PAMELA data on positron fraction, in a single-component scenario adopting different propagation setups; we find that the model is not adequate to reproduce the two datasets, so the evidence of an extra primary component of electrons and positrons is strengthened. Instead, a double component scenario computed in a Kraichnan-like diffusion setup (which is suggested by B/C and $\bar{p}$ data) gives a satisfactory fit of all exisiting measurements. We confirm that nearby pulsars are good source candidates for the required $e^\pm$ extra-component and we show that the predicted CRE anisotropy in our scenario is compatible with Fermi-LAT recently published constraints.
\end{abstract}

\maketitle



\section{Introduction}

Last year the Fermi-LAT Collaboration published the $e^+ + e^-$ spectrum in the energy range between 20 GeV and 1 TeV, measured during the first six months of the Fermi mission\cite{Abdo:2009zk}. That result came in the middle of an interesting debate which arose after that ATIC and PAMELA collaborations reported some anomalous results: ATIC observed a bump in the $e^+ + e^-$ spectrum at around 600 GeV, while PAMELA found the positron fraction $e^+/(e^+ + e^-)$ to increase with energy above 10 GeV: these features are hardly compatible with the standard scenario in which CR electrons are accelerated in supernova remnants (SNRs) and positrons are predominantly of secondary origin, and were interpreted by many authors as a possible signature of decay or annihilation of Dark Matter (DM) particles, even if a more conventional interpretation in terms of astrophysical sources (namely pulsars) was also considered. Fermi-LAT spectrum does not display the feature seen by ATIC, being compatible with a single power law; the absence of that feature was later confirmed by the H.E.S.S. Cherenkov telescope whose spectrum below 1 TeV is in agreement with Fermi-LAT's. Soon after that measurement was published, some of us, with other members of Fermi collaboration, showed that an interpretation of the $e^+ + e^-$ spectrum is possible within a conventional model in which SNRs are the only primary sources of CREs\cite{CRE_interpretation1}. Below 20 GeV, however, we found the predictions of that model to be in tension with pre-Fermi data; furthermore, the positron fraction measured by PAMELA was not reproduced within that framework. We therefore proposed, in the same paper, a scenario which invoked the presence of an extra component of $e^- + e^+$: it was argued that such extra-component can naturally be produced by near middle-aged pulsars. An alternative interpretation based on the annihilation of DM particles was also discussed but considered disfavoured. Other possible origins of the required $e^+ + e^-$ extra-component were proposed by other Authors: e.g. enhanced secondary production in standard SNRs\cite{Blasi:2009hv} or an inhomogeneous distribution of sources in the Galaxy\cite{Shaviv:2009bu}. Recently, the Fermi-LAT Collaboration released a new measurement of the CRE spectrum based on one year data. The observed spectrum extends down to 7 GeV\cite{Ackermann:2010ij} and is confirmed to be compatible with a single power-law with spectral index $3.08 \pm 0.05$. Hints of a deviation from a pure power-law behavior, already found in the six month data, are still present in the updated spectrum, whose low energy part is compatible with pre-Fermi measurement and not reproduced by the models we proposed in Ref. 2: this consideration calls for some revisions of our scenario. Moreover, after the release of PAMELA antiproton data some of us used (see Ref. 6) the recently developed {\tt DRAGON}~\footnote{Code available at http://www.desy.de/$\sim$maccione/DRAGON/}  package\cite{Evoli:2008dv} to perform a combined maximum likelihood analysis on B/C and antiproton data and found that the ``conventional model'', based on the assumption of the existence a Kolmogorov-like turbulence in the ISM, is not the most adequate to describe the measurements: we therefore proposed a new model based on a Kraichnan-like turbulence. This result gives another reason to revise the models described in Ref. 2. 

\section{The propagation setups}\label{sec:propagation}

\begin{figure}[tbph]
  \centering  
  \subfigure[]
  {
   \includegraphics[scale=0.35]{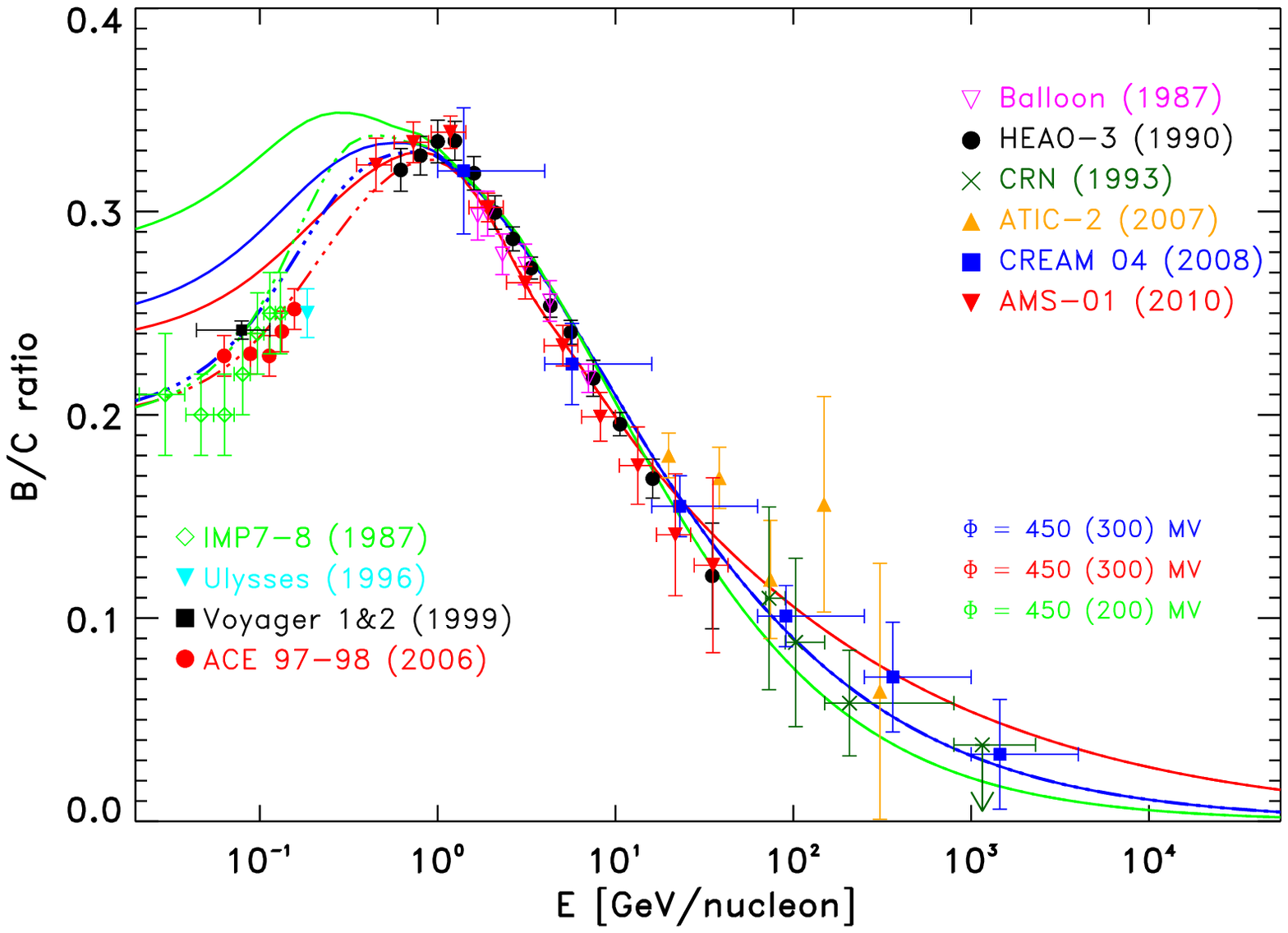}
   \label{fig:BoverC}
   }
   \subfigure[]
   {
   \includegraphics[scale=0.35]{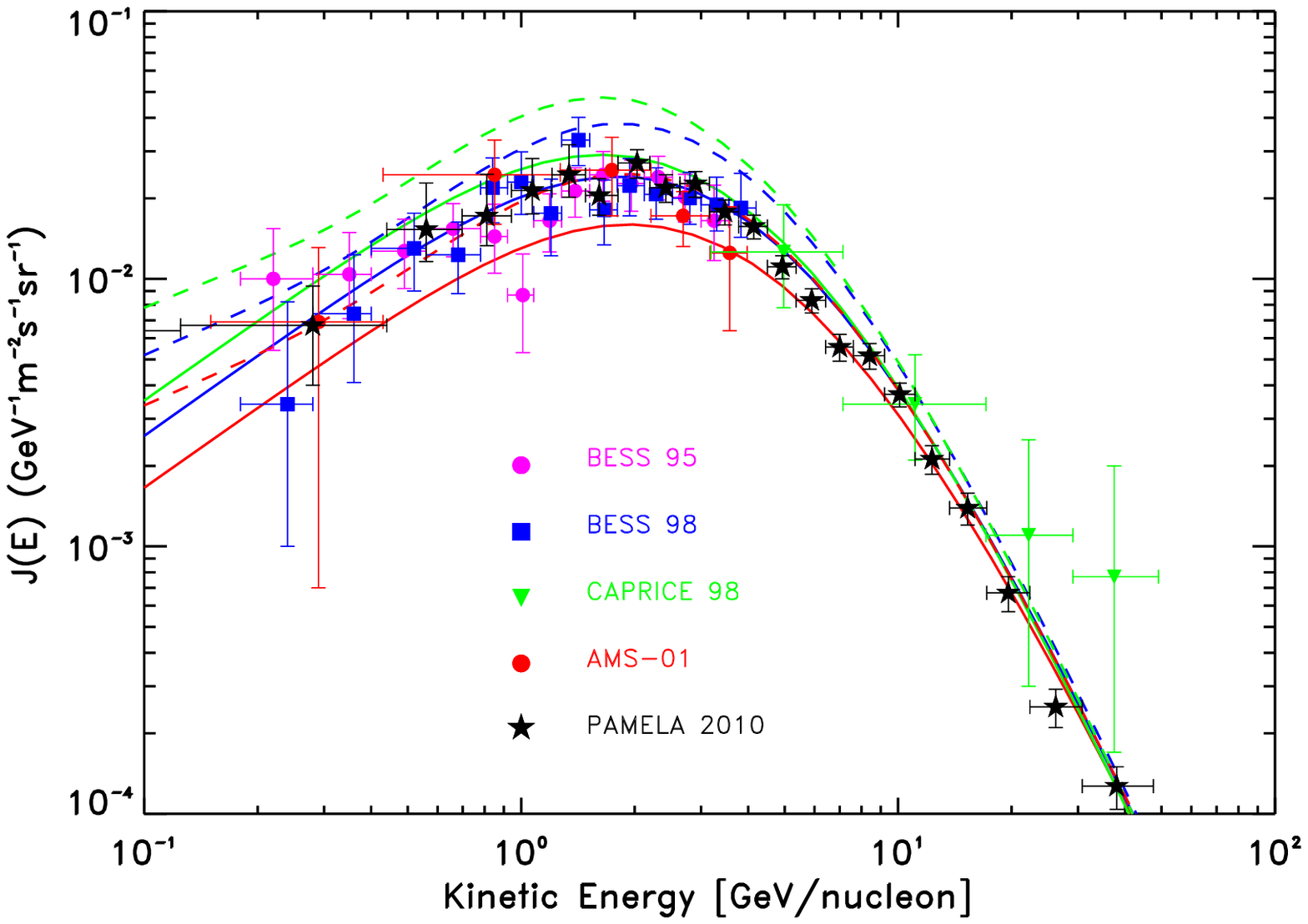}
   \label{fig:antip_p}
   }
\caption{\footnotesize Panel a) Predicted B/C compared with data. Red lines: Kolmogorov model  (KOL);  blue: Kraichnan model (KRA); green: plain diffusion (PD). Solid lines: $\Phi = 450~{\rm MV}$; triple-dotted lines: $\Phi = 300~{\rm MV}$. Panel b) Theoretical ${\bar p}$ spectra are compared with data; solid lines: $\Phi = 550~{\rm MV}$; dashed lines: LIS.}
\end{figure}

%
The CR propagation in the Galaxy is described by a well known diffusion-loss equation which can be solved analitically, under simplifying assumptions, or numerically, making use of  {\tt GALPROP} or  {\tt DRAGON} packages. This equation includes several free parameters which need to be tuned by comparison with data: $D_{0}$ and $\delta$, i.e. the normalization and energy dependence of the diffusion coefficient, the Alfv\'en velocity $v_{A}$ which parametrizes the level of reacceleration, the height of the Galactic diffusion region $z_{h}$, and the injection index of the CR species $\gamma_{p}$.  Moreover, when considering data below a few GeV/n also the modulation due to solar activity plays a significant role and must be taken into account. The first step of our work, before starting to investigate the $e^+ + e^-$ spectrum, consists in fixing the propagation parameters, i.e. $D_{0}$ and $\delta$, and the level of reacceleration. The datasets which constrain these data are secondary-to-primary ratios, in particular B/C and ${\bar p}/p$. In the following we will consider three propagation models, which are defined as follows. 1) The PD model is a plain diffusion one, in which we tried to reproduce CR data with no reacceleration, hence setting $v_{A} = 0$. In this model $\delta = 0.6$. 2) The KOL model is the so called ``conventional model'', built assuming a Kolmogorov-like turbulence, which fixes $\delta = 1/3$; in this model the reacceleration is quite high: $v_{A} = 30\, {\rm km/s}$. 3) The KRA model assumes a Kraichnan spectrum for the galactic turbulent magnetic field, hence setting $\delta = 1/2$. Reacceleration is lower: $v_{A} = 15\, {\rm km/s}$. At low energy a modified behaviour of the diffusion coefficient is adopted to reproduce the peak at 1 GeV in B/C data. This model was presented in Ref. 6 and obtained as the result of a combined maximum likelihood analysis based on B/C and ${\bar p}/p$ measurements. The comparison between data and model predictions for B/C and ${\bar p}/p$ can be seen in Fig. \ref{fig:BoverC} and Fig. \ref{fig:antip_p}. We refer to Ref. 6 for the complete list of parameters we chose for the 3 models.

\section{Modeling the CRE spectrum with a single Galactic component}
\label{sec:single_comp}

\begin{figure}[tbph]
  \centering
   \subfigure[]
  {
   \includegraphics[scale=0.35]{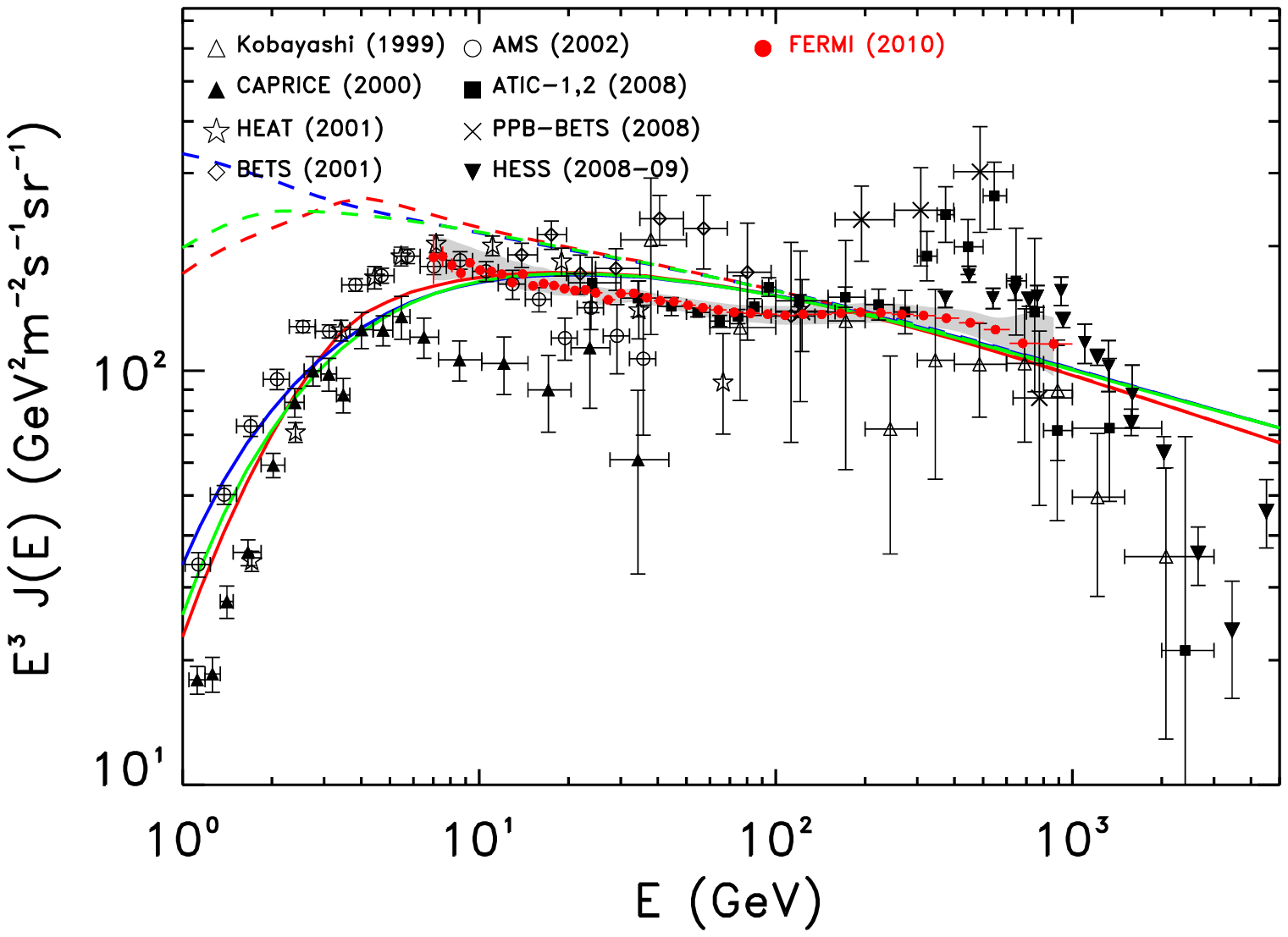}
   \label{fig:singlecomponent}
   }
     \subfigure[]
  {
   \includegraphics[scale=0.35]{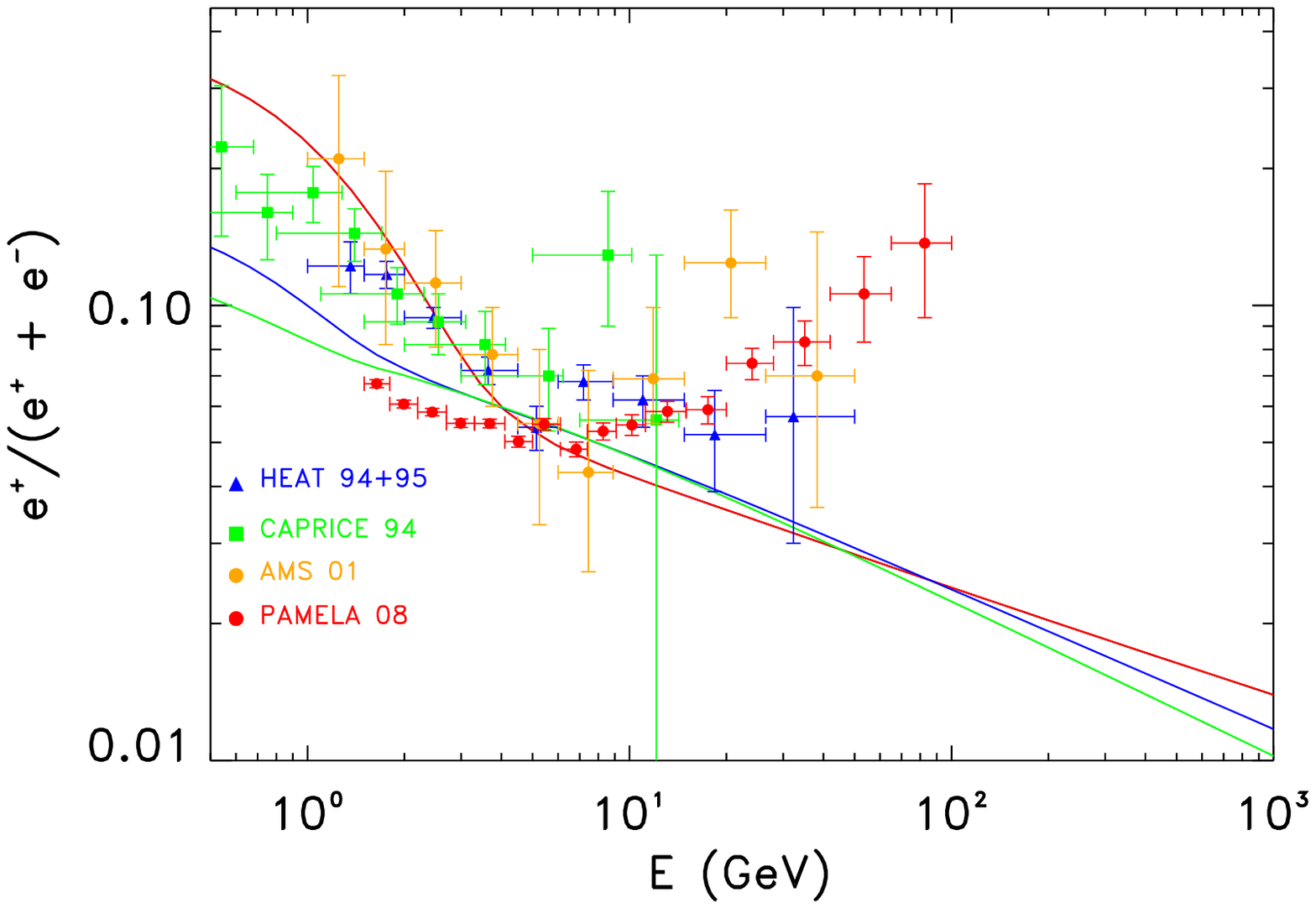}
   \label{fig:singlecomponent_posratio}
   }
\caption{\footnotesize Panel a): the $e^- + e^+ $ spectrum for the KOL (red), KRA (blue) and PD (green) diffusion setups. The electron source spectral indexes are 1.60/2.50 below/above 4 GeV for the KOL model and 2.00/2.43 and 2.0/2.40 below/above 2 GeV for the KRA  and PD models respectively. Panel b):  $e^+/(e^- + e^+ )$.  $\Phi = 550~{\rm MV}$.}
\end{figure}

We start our analysis by trying to interpret CRE spectrum with single-component models.  We evaluated these model with {\tt DRAGON} numerical package\cite{DiBernardo:2009ku} and verified that our results are reproduced by {\tt GALPROP}\cite{galprop,GALPROPweb} under the same physical conditions. We consider the three diffusion setups (KOL, KRA and PD) discussed in the previous section and for each of them we tune the free parameters involved in the calculation against Fermi-LAT data. We find that a reasonable fit of Fermi-LAT spectrum can be obtained normalizing the models to data at $\simeq 10$ GeV and adopting the following injection indexes: $2.50$ in the KRA model, $2.43$ for the KOL, $2.40$ for the PD (see Fig. \ref{fig:singlecomponent}). It is worth noticing that KOL model requires a sharp spectral break (i.e. an injection index of $1.60$  below $4$ GeV) to avoid an anomalous behavior which would otherwise arise in the propagated spectrum: the combined effect of reacceleration and energy losses creates a pronounced bump in the unmodulated spectrum at low energy, which is only partially smeared out by solar modulation. This effect is smaller in KRA model and absent in PD setup, so in these cases a much softer break is required: $2.0$ instead of $1.60$ is the index below $2$ GeV. However, the main drawback of a single component approach, concerning Fermi-LAT data, is that -- no matter what propagation model is adopted -- it is impossible to reproduce all the features revealed in the CRE spectrum, in particular the flattening observed at around 20 GeV (which was also recently found by PAMELA \cite{Adriani_talk}) and the softening at $\sim 500$ GeV. Concerning the positron fraction, the results can be seen in Fig.~\ref{fig:singlecomponent_posratio}. Clearly, below 10 GeV the $e^+/(e^- + e^+)$  measured by PAMELA can be reproduced in the KRA and PD models while the fit is unsatisfactory for the KOL: again, low reacceleration models seem to provide a better description of low energy data. None of the single component realizations, however,  can reproduce the rising behaviour observed by PAMELA above 10 GeV.

\section{Two components models}
\label{sec:toy_model}

\begin{figure}[tbp]
  \centering 
  \includegraphics[scale=0.45]{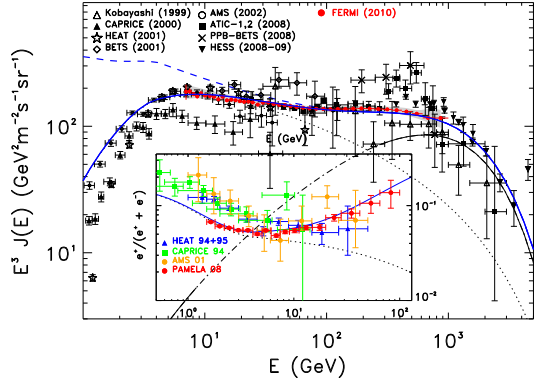}
  \label{fig:extra_component_fit}
  \caption{\footnotesize The $e^- + e^+$ total spectrum and positron fraction (in the box) for our two-component model. Dotted line: propagated standard component with injection slope $\gamma_{e^-} = 2.00/2.65$ above/below $4~\GeV$ and $E^{e^-}_{\rm cut} = 3~\TeV$; dot-dashed line: $e^\pm$ component with  $\gamma_{e^\pm} =  1.5$ and $E^{e^\pm}_{\rm cut} = 1.4~\TeV$. Blue solid line: modulated total spectrum ($\Phi = 550$ MV). Blue dashed: LIS total.}
\end{figure}

In the following we try to reproduce all data in a two-component scenario. The first component (standard) consists of electrons accelerated in SNRs and a secondary contribution of $e^-$ and $e^+$ originated by the interaction of the hadronic part of CRs with interstellar gas. The second one (extra-component) is made of $e^-$ and $e^+$ injected in the ISM with a common spectrum of this kind: 
$Q_{e^\pm}(E) = Q_{0}\left(\frac{E}{E_{0}}\right)^{- \gamma_{e^\pm}}e^{-E/E^{e^\pm}_{\rm cut}}$,
where the injection index is harder: $\gamma_{e^\pm} <  2$. We also assume that both source classes have the same continuous spatial distribution. The normalization of the two components is performed in two steps. 
1) We tune the standard component to reproduce both the $e^- + e^+$ spectrum measured by Fermi-LAT and the  $e^+/(e^- + e^+)$ measured by PAMELA below 20 GeV, where the effect of the extra component is supposed to be negligible. Remarkably, this is possible only if we use propagation setups with low reacceleration, namely either the KRA or the PD, because the low-energy part of PAMELA positron ratio can't be reproduced in KOL setup. Since the KRA\cite{DiBernardo:2009ku} also provides the best combined fit of B/C and antiproton data (see Sec.~\ref{sec:propagation}), we will stick to this model from now on. The required source spectral slopes for the electron standard component is  $\gamma_{e^-} = 2.00/2.65$  below/above 4 GeV for this propagation model. From Fig. 3 the reader can see as such a set-up allows a remarkably good fit of Fermi-LAT as well as other experimental data.
Such index is quite steep if compared to theoretical predictions regarding Fermi acceleration mechanism, but we remind the reader that we modeled the standard component in the approximation of a cylindrically symmetric source distribution, which may be less realistic for high energy electrons where the local distribution is relevant.  Accounting for the spiral arm distribution of SNRs may result in a different requirement for the injection index. Indeed, being the Sun in the so-called ``local spur'' situated in a interarm region, the average distance from SNRs is larger than in the smooth case: as a consequence, a harder injection spectrum may be required to compensate for the larger energy losses and reproduce the observed spectrum. Clearly, in the absence of the extra $e^\pm $component, high energy CRE and positron fraction data would completely be missed (see dotted line in Fig.~3).
2) We tune the extra-component to reproduce Fermi-LAT and H.E.S.S. high energy CRE data. We find here that this is possible by taking $\gamma_{e^\pm} = 1.5$ and  $E_{\rm cut} = 1.0 \div 1.5 ~\TeV$ (see Fig.~3). This is similar to what done in Ref. 2 but in that paper a KOL diffusion was used (so low energy positron data were not reproduced in that case).  It is interesting that preliminary $e^-$ spectrum measured by PAMELA \cite{Adriani_talk} is also nicely reproduced by our extra-component models: above 100 GeV this spectrum is softer than the $e^- + e^+$ measured by Fermi-LAT by the exact amount which is required to leave room the $e^+$ extra-component. 

\section{The role of astrophysical nearby sources}\label{sec:discrete_sources}

\begin{figure}[tbp]
\begin{center}
 \centering
  \subfigure[]
 {
  \includegraphics[scale=0.35]{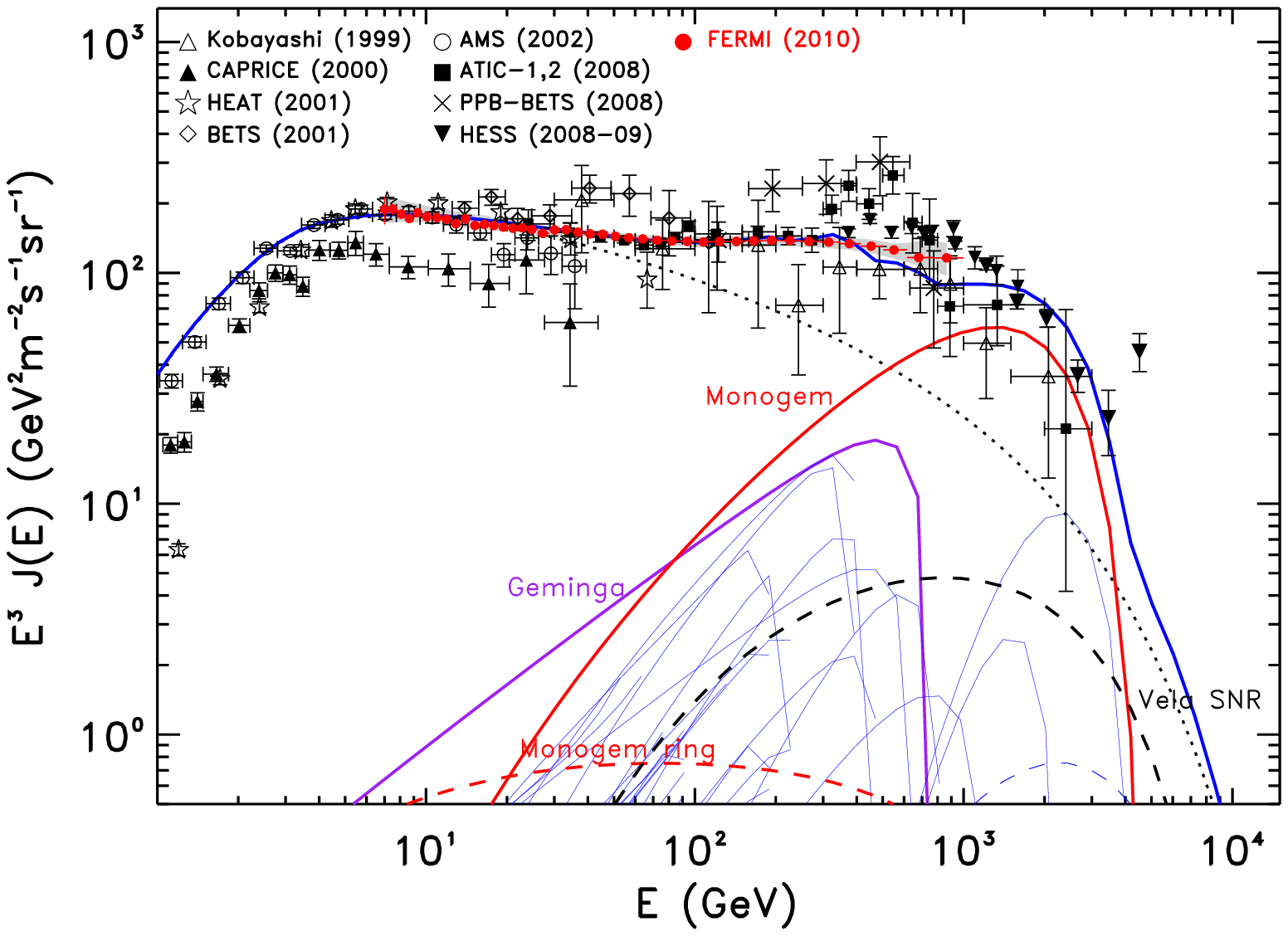}
  \label{fig:Hybrid_1_spectrum}
  }
 \subfigure[]
 {
 \includegraphics[scale=0.35]{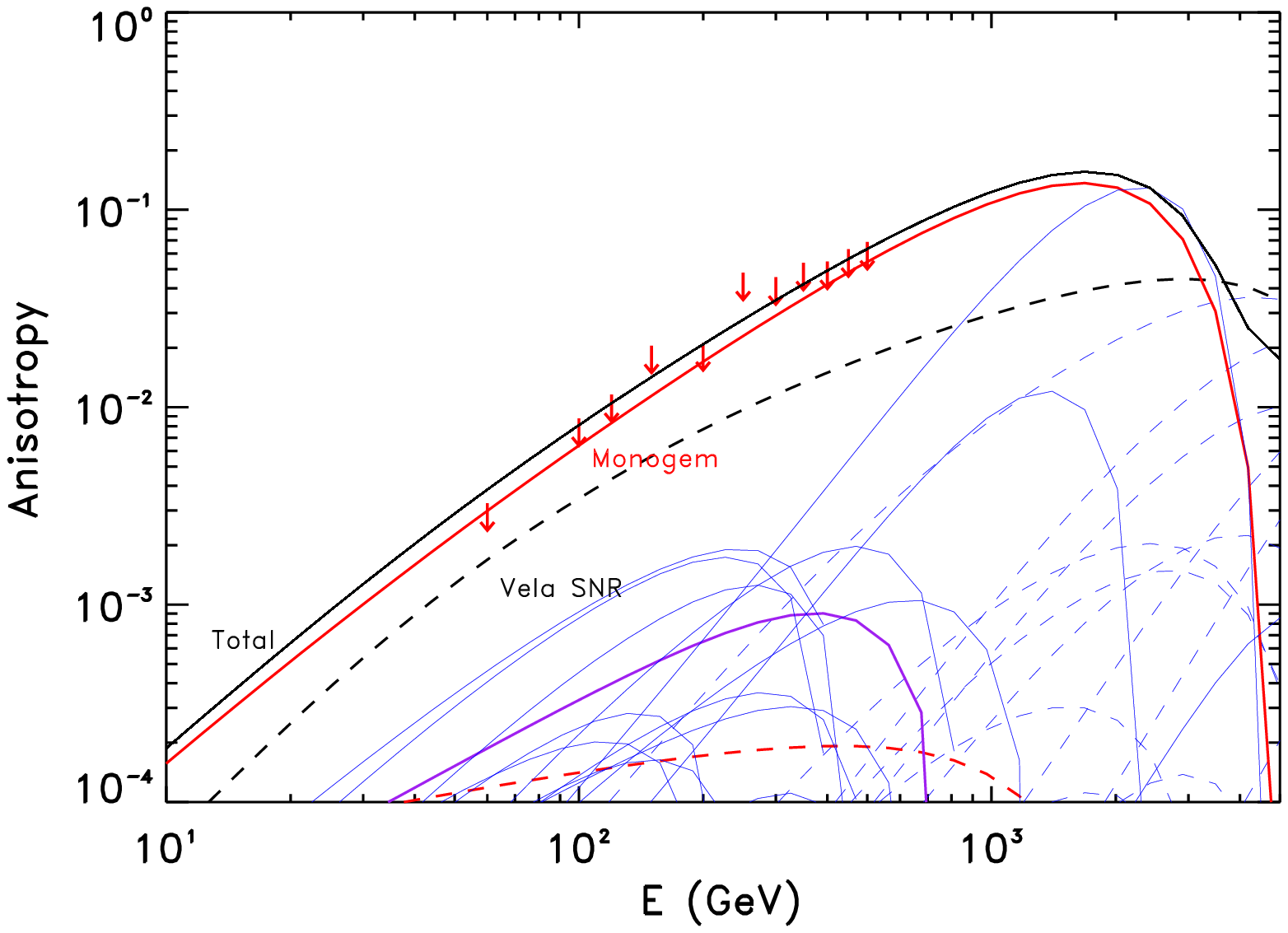}
  \label{fig:Hybrid_1_aniso}
  }    
\caption{\footnotesize Standard component + nearby pulsars and SNRs. Panel a): $e^+ + e^-$ spectrum. Energy release by each SNR: $2 \times 10^{47}\,{\rm erg}$ .  Pulsar efficiency: $\simeq 30\%$. $\Phi = 500 MV$. Panel b): the integrated expected anisotropy, as a function of minimum energy, is compared to Fermi-LAT the upper limits \cite{Ackermann:2010ip}.}
\end{center}
\end{figure}

The nature of the extra-component of primary electrons and positrons that we invoked in the previous section is an intriguing matter of debate, and the possible scenarios include both an exotic explanation (involving annihilation or decay of Particle Dark Matter) or a purely astrophysical interpretation. Here we concentrate on the second possibility.  Differently from the previous section, we treat the extra component as originating from a discrete collection of sources; we then treat electron and positron propagation from those sources to the Solar System by solving analytically the diffusion-loss equation similarly to what done in Ref. 2. The large scale (standard) component is instead modeled with {\tt DRAGON} as done in the previous section; for consistency, we treat analytical and numerical propagation under the same physical conditions. The propagation setup adopted is the KRA one.  There are basically two classes of objects we consider: SNRs (as sources of {\it electrons only}) and pulsars (as sources of {\it electron+positron pairs}). SNRs are the well known natural candidates as CR accelerators, and also the conventional component is expected to originate from objects of this kind; considering individual SNRs in the nearby ISM is important due to the fact that at high energy (above some hundred GeV) diffuse models don't give a proper description of reality since the loss length due to synchrotron and IC becomes comparable to the average SNR mutual distance so that only few sources within few hundred parsecs are expected to dominate. Pulsars, instead, are very interesting candidates as primary sources of both electrons and positrons. They are extreme astrophysical environments that release very large amounts of energy ($\sim 10^{52} \div 10^{54}$ erg) during their lifetime. The $e^- + e^+$ pairs are expected to be produced in the magnetosphere by the interaction of curvature photons with the pulsar magnetic field; those particles are then expected to be accelerated at the termination shock of the pulsar wind nebula (PWN). The possibility that electrons and positrons from nearby pulsars can dominate the high energy tail of the CRE spectrum and explain the rising behavior of the positron fraction was already proposed in Ref. 13 and studied in more recent papers (e.g.~ Refs 14 and 15). Similarly to the approach taken in Ref. 2, we model the emission from SNRs and pulsars as a point-like burst; for pulsars we also introduce a time delay with respect to the birth of the object: such a delay is motivated by the fact that electrons and positrons are expected to be trapped in the PWN until it merges with the ISM. We assume that for both SNRs and pulsars particles are injected with a power law spectrum up to an exponential cutoff; the injection index is however set in a different way for the two classes of objects. We considered all observed SNRs within 2 kpc as taken from the Green catalogue \cite{Green:2009qf} and the pulsars within 2 kpc from Earth taken from the ATNF catalogue \cite{Manchester:05}. We verified that more distant objects give a negligible contribution. In Fig. \ref{fig:Hybrid_1_spectrum} we represent the CRE spectrum obtained for a reasonable combination of  parameters, namely: for SNRs:  spectral index $\gamma_{e^{- \,{\rm SNR}}} = 2.2$, cutoff energy $E^{\rm SNR}_{\rm cut} = 2~\TeV$,  electron energy release per SN  $E^{\rm SNR} =  2 \times 10^{47}\,{\rm erg}$;  for pulsars:  $\gamma_{e^{\pm}} = 1.5$,  cutoff energy $E_{\rm cut} = 1~\TeV$, efficiency $\eta_{e^\pm} \simeq 30\%$. We see from Fig. \ref {fig:Hybrid_1_spectrum} that the main contribution comes from Monogem pulsar; this is due to the proximity of this source and to the introduction of the delay between the pulsar birth and the actual injection of the pairs in the ISM; without considering such delay, Monogem contribution would decrease significantly and other nearby young pulsar, such as Vela, would be the most important sources in the high energy region; instead, in our framework electrons and positrons from Vela are expected to be still trapped in the surrounding nebula and therefore not yet observed in the CRE spectrum. It is important to check if these results are compatible with recently published upper limits on the anisotropy in $e^- + e^+$ flux \cite{Ackermann:2010ip} . Our prediction is plotted in Fig. \ref{fig:Hybrid_1_aniso}. The important result is that our scenario is not excluded by anisotropy measurements; the reader may notice that Monogem pulsar (red solid line) and Vela SNR (black dashed line) contribute most to the total anisotropy (the black solid line). However,  the total expected anisotropy is very close to the measured upper limit, so a future detection at level $\sim 1\%$ at $\sim 1$ TeV towards the portion of the sky where Vela and Monogem are located is to be expected.

\section{Conclusions}

The spectacular data on CR electrons and positrons, together with measurements on light nuclei and antiprotons, suggest that a double-component approach to the leptonic part of CRs computed in the framework of a Kraichnan-like turbulence provides a good self-consistent scenario which satisfactorily reproduces all existing observations. In this picture, in addition to the conventional component accelerated in SNRs, a contribution from pulsars, as emitters of $e^- + e^+$, permits to correctly fit both the features revealed by Fermi-LAT in the CRE spectrum and the $e^-/(e^+ + e^-)$ measured by PAMELA. The expected anisotropy in the direction of the most prominent CRE candidate source, Monogem pulsar, is compatible with the present upper limit just released by Fermi-LAT collaboration and may be detectable in a few years.


\begin{thebibliography}{9}

\footnotesize

 \bibitem{Abdo:2009zk}
  A.~A.~Abdo {\it et al.}  [The Fermi LAT Collaboration],
  Phys.\ Rev.\ Lett.\  {102} (2009) 181101
  
  \bibitem{CRE_interpretation1}
  D.~Grasso {\it et al.}  [FERMI-LAT Collaboration],
  Astropart.\ Phys.\  {32} (2009) 140
  
 \bibitem{Blasi:2009hv}
  P.~Blasi,
  Phys.\ Rev.\ Lett.\  {103} (2009) 051104
  
 \bibitem{Shaviv:2009bu}
  N.~J.~Shaviv, E.~Nakar and T.~Piran,
  Phys.\ Rev.\ Lett.\  {103} (2009) 111302

  \bibitem{Ackermann:2010ij}
  M.~Ackermann {\it et al.}  [Fermi LAT Collaboration],
  arXiv:1008.3999 [astro-ph.HE], accepted for publication in Phys.\ Rev.\ D
  
 \bibitem{DiBernardo:2009ku}
  G.~Di Bernardo, C.~Evoli, D.~Gaggero, D.~Grasso and L.~Maccione,
  arXiv:0909.4548 [astro-ph.HE];    Astropart.\ Phys., in press\  
  doi:10.1016/j.astropartphys.2010.08.006 .

  \bibitem{Evoli:2008dv}
  C.~Evoli, D.~Gaggero, D.~Grasso and L.~Maccione,
  {\it JCAP} {0810} (2008) 018\ 

 \bibitem{DiBernardo:2010}
  G.~Di Bernardo, C.~Evoli, D.~Gaggero, D.~Grasso, L.~Maccione and M.N.~Mazziotta
  arXiv:1010.0174 [astro-ph.HE]; submitted to Astropart.\ Phys.
  
  \bibitem{galprop}
  A.~W.~Strong, I.~V.~Moskalenko, T.~A.~Porter, G.~Johannesson, E.~Orlando and S.~W.~Digel,
  arXiv:0907.0559 [astro-ph.HE].

  \bibitem{GALPROPweb}
  A.~E.~Vladimirov {\it et al.},
  arXiv:1008.3642 [astro-ph.HE].
 
 \bibitem{Adriani_talk}
 O.~ Adriani, talk at ICHEP conference 2010, Paris (http://www.ichep2010.fr/)

  \bibitem{Ackermann:2010ip}
  M.~Ackermann  {\it et al.} [Fermi-LAT collaboration]
  arXiv:1008.5119 [astro-ph.HE]; accepted for publication in Phys.\ Rev.\ D.
  
 \bibitem{Aharonian:1995zz}
  F.~A.~Aharonian, A.~M.~Atoyan and H.~J.~Volk,
  Astron.\ Astrophys.\  {294} (1995) L41.
    
 \bibitem{Hooper:2008kg}
 D.~Hooper, P.~Blasi and P.~D.~Serpico,
 JCAP\  {0901} (2009) 025.
 
 \bibitem{Profumo:2008ms}
  S.~Profumo,
  arXiv:0812.4457 [astro-ph].
 
 \bibitem{Green:2009qf}
  D.~A.~Green,
   Bull.\ Astron.\ Soc.\ Ind., 37 (2009)  45;  
  arXiv:0905.3699 [astro-ph.HE].

 \bibitem{Manchester:05} 
 R.~N.~Manchester, G.~B.~Hobbs, A.~Teoh, \& M.~Hobbs, ApJ\ 129 (2005) 1993


\end{thebibliography}
\end{document}